# An Efficient and Compact Excitation of Surface Plasmons Using a Si Gable Tip

[1]Arnab Dewanjee , [1,2]M. Z. Alam, [1] J. Stewart Aitchison [1]Mo. Mojahedi
[1]Department of Electrical and Computer Engineering,
University of Toronto, 40 King's College road, Ontario-M5S3G4, Canada
Now: California Institute of Technology, Pasadena, California, USA
Author email: arnab.dewanjee@mail.utoronto.ca

**Abstract:** *We propose a novel technique to efficiently excite a surface plasmon polariton (SPP) mode at a gold-glass interface by exploiting the near field excitation of an engineered high index (silicon) gabled tip at the telecom wavelengths. The proposed structure can theoretically convert 49% of the input optical power to a SPP mode. The device is compact, it will facilitate the on-chip excitation of the SPP, its fabrication is compatible with the standard Si fabrication processes, and as such, it is expected to be useful in the design of future integrated sensors.*

Surface plasmon polaritons (SPP) are the propagation of electromagnetic wave coupled with a collective electron oscillations and confined along a metal-dielectric interface [1]. Due to the subwavelength nature and high spatial confinement the SPP mode sizes can reach beyond diffraction limit. Hence, for the purpose of confining light in the smallest possible dimensions SPPs are naturally a very convenient choice [1]. Thus SPPs have found applications in many emerging areas of nanophotonics such as integrated optics [3,4], field enhancement [5], sensing [6] and imaging [7,8]. Advancements have been made to improve the directionality and controllability of an excited SPP mode [9-13]. Similarly an efficient excitation of the SPP mode from the source is of equal importance for nanophotonic applications since SPP modes are inherently lossy due to the resistive loss of the metal. There have been a number of SPP excitation schemes proposed in the literature. The most commonly used SPP excitation schemes are prism coupling in the Otto or Kretschmann configuration [14] and metallic grating coupling [15-17]. Tianran Liu *et al.* demonstrated a high efficiency dual layer grating SPP coupler for a gold-air interface [17]. Pors *et al.* proposed and demonstrated 1D and 2D gap resonators for directional coupling of SPP modes [16]. Such couplers are highly efficient but require the out of plane bulky and angle sensitive coupling set up and perform well at narrow input beam cross sections. Therefore, such SPP coupling schemes are not suitable for compact, portable and integrated SPP excitation. For more compact coupling of SPP, scattering of light by transmission through a metallic nano-hole has also been demonstrated [11-13,18]. Jing yang *et al.* used an asymmetric optical slot nano antenna pair to directionally excite SPP modes on a gold-air interface [11]. Lin *et al.* demonstrated polarization controlled tunable directional coupling of SPP modes using polarization sensitive apertures in a gold film [12]. Yao *et al.* demonstrated directional SPP excitation using single element nano groves on a gold film [13]. In all these cases the nano-hole SPP excitation scheme suffers from low excitation efficiency even when efficiency is calculated by normalizing the power in the SPP mode by the incident power on the hole. The power incident on the geometric area of the hole is typically only a fraction of the total power carried by the incident beam. Therefore, SPP coupling efficiency numbers are even lower when the coupled SPP mode power is normalized to the total power in the beam. One possible solution to increasing the overall efficiency is to tightly focus the beam so that maximum power is incident on the aperture of the nanohole. But, due to the diffraction limit, light cannot be focused tighter than approximately half of the wavelength while the SPP exciting nano apertures have much smaller dimensions. However, Lalanne *et al.* have shown that the coupling efficiency of an SPP mode by scattering from a metal nano-hole gradually decreases with increasing width of the hole [19]. Therefore increasing the hole width to capture more power from the incident light does not improve the SPP coupling efficiency. Additionally, the SPP excitation efficiency further decreases with the increase in wavelength [19]. Consequently, the SPP excitation efficiency in the telecom wavelength (1550 nm) is significantly lower than that in the visible region with the popular compact coupling techniques. To be compatible to the large variety of applications in the telecom wavelength regime, a compact and efficient method to excite SPP at the 1550 nm wavelength is very important. Mueller *et al.* have demonstrated fishbone structured plasmonic nano antenna array for SPP coupling at 1550 nm; but with a very low coupling efficiency (<1%) [11]. Li *et al.* demonstrated SPP coupling to a silver nanowire at 1550 nm using a nano-tapered tip of fiber where the coupling efficiency is low and the setup demands extreme precision [20]. In this study we present a new technique with a silicon (Si) nano-tip structure with properly chosen tip angle to excite surface plasmon polariton at a metal-dielectric interface in the telecom wavelength (1550 nm) regime which has been inspired by some earlier work on near field coupling techniques of light.

In their seminal paper Reddick *et al.* demonstrated a near field photon tunneling scanning microscope (PSTM) capable of coupling evanescent light at an air-dielectric (glass) interface excited by total internal reflection [21]. In a similar fashion, an SPP mode can also be excited by total internal refraction using the Otto or Kretschmann configuration, as mentioned earlier. SPP modes excited by total internal refraction can also be imaged using PSTM by bringing the dielectric tip sufficiently close to the metal dielectric interface [22]. In both [21] and [22] tunneling of photons through evanescently decaying field at the boundary of a metal dielectric interface has been utilized to couple to a propagating mode inside a dielectric tip structure placed in the near field. Metal coated dielectric tips have also been used to excite SPP at a metal/dielectric interface with relatively low excitation efficiency (~ 0.3%) [23]. Here we investigate the use of a pure dielectric tip structure to excite the SPP mode (at 1550 nm wavelength) at the metal dielectric interface through nearfield coupling by tunneling of photon. The excitation efficiency of the SPP mode at the metal/dielectric interface at the near field of the tip depends on several parameters such as the shape of the tip, the distance of the tip from the interface and the refractive index of the tip as well as the medium surrounding the tip. In this paper we investigate and carefully design a novel high-index dielectric gabled tip to directionally excite an SPP mode at a gold-glass interface at the telecom wavelength. Since our goal is to efficiently excite SPP in the 1550 nm wavelength Si is the convenient choice for the low loss high index material for the the inverted tip.

Figure 1(a) shows the 2D schematic of the proposed setup for the efficient excitation of an SPP mode. The base (red in figure 1-a) of the Si tip collects the incident light from the beam with a large cross-section and concentrates the energy delivered by the beam at the tip without leakage. The concentrated light from the tip then excites two propagating SPP modes towards the right and the left (as shown by the blue arrows in figure 1-a) at the gold-glass interface by means of near field coupling. A commercial-grade simulator based on the finite-difference time-domain method was used in our work to simulate, analyze and optimize the performance of the proposed device[24]. Figure 1 (b) shows the simulated power distribution of the SPP coupling setup shown in figure 1 (a). As we will establish later in this paper that the slope of the Si tip plays important role in an efficient coupling of the SPP, the tip height can be adjusted to increase the base size (keeping the slope constant) to facilitate the capture of a larger beam cross section with a prior knowledge of the beam size. Hence, unlike most of the cases in the literature, the SPP excitation efficiency for the proposed technique can be calculated with respect to the total energy in the incident beam from the source rather than a portion of the beam cross sectional area determined by a nanoscale aperture of the excitation geometry. Additionally, the capability to capture large beam cross-sections the proposed SPP coupling scheme helps get rid of the necessity of additional optical components to reduce the beam size to the size of a few microns.

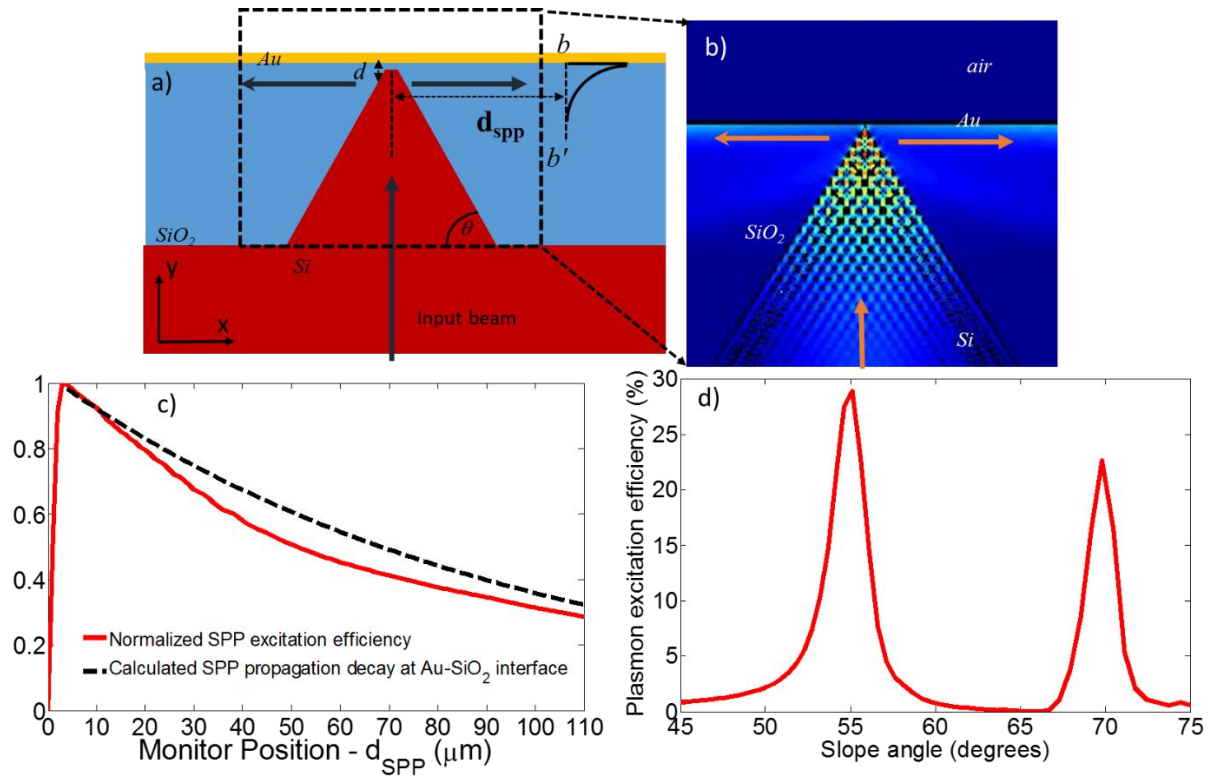

Figure: 1. a) Schematic of the SPP excitation setup by an inverted Si tip on a Si substrate. b) Simulated power profile of the SPP excitation system in (a). c) Normalized SPP coupling efficiency as a function of probing distance from the inverted Si tip - $d_{SPP}$ (red). For ease of comparison the calculated SPP decay profile is also presented (black dotted). d) SPP excitation efficiency as a function of the slope angle of the inverted Si tip.

To achieve the goal of a high SPP coupling efficiency we have investigate the determining factors of the near-field SPP mode coupling from a high index dielectric tip to a metal-dielectric interface. However, before delving into the results of such investigations it is necessary to establish the proper definition of the SPP excitation efficiency in light of the proper probing distance from the Si tip since the excited SPP mode exponentially decays as it propagates due to loss in the metal. The red curve in figure 1(c) plots the normalized power in the excited SPP mode (by the Si tip) at 1550 nm wavelength as a function of the probing distance ($d_{SPP}$ in figure 1-a) from the tip. The black dashed curve (figure 1-c) shows the theoretically calculated exponential decay profile of a propagating SPP mode at a gold-glass interface. As can be observed from the two curves in figure 1(c), the normalized power in the excited SPP mode increases rapidly as $d_{SPP}$ increases from 0 to 4 μm and then exponentially decays following the black dashed curve with further increase of the probing distance ($d_{SPP}$). This is an indication of the fact that light scattered from the inverted Si tip experiences near field interaction with the nearby metal-dielectric (Au-SiO$_2$) interface and gradually evolves into a coupled light-electron oscillation that starts propagating along the Au-SiO$_2$ interface. This evolution takes place in the region up to a 4 μm distance and from there the propagating mode attains the properties of the SPP mode. We consider this $d_{SPP}$ = 4 μm to be the SPP excitation distance and calculate all the SPP excitation efficiencies at this distance from the inverted Si tip in our presented simulation results in the rest of the paper.

At this point we discuss the effect of various geometrical parameters of the proposed device on the SPP coupling efficiency. Since exciting an SPP mode necessitates the generation of sufficiently matched momentum along the Au-SiO$_2$ interface, the shape of the tip plays an important role in excitation process. Additionally, the distance between the tip and the interface also affects the SPP excitation efficiency. First we look into the effect of slope angle ($\theta$), in other terms, the tip angle, on the efficiency of the SPP excitation since achieving an arbitrary slope is the most challenging part of the fabrication process of the geometry in Figure 1(a). Figure 1(d) plots the SPP excitation efficiency (considering one direction i.e. the right side of the Si tip in figure 1-b) against the slope angle ($\theta$) of the inverted Si tip. In figure 1(d), there are two excitation peaks at $55^0$ and $70^0$ and the stronger peak is at $55^0$. This SPP excitation efficiency peak at $55^0$ for the slope angle is a very convenient outcome since the KOH wet etch of a <100> Si wafer etches at a slope angle of $54.7^0$, thus making the fabrication of the inverted Si tip optimized for SPP excitation quite feasible and inexpensive. From this point in the paper, for all the calculations, $54.7^0$ is chosen as the slope angle of the inverted Si tip. To optimize the distance (*d* in figure 1) between the tip and the Au-SiO$_2$ interface, we consider two cases of the tip geometry- a sharp tip and then a 500 nm wide tip. The dimensions of the tip of the fabricated device presented later in this paper motivated us to consider of a flat tip and compare the SPP coupling efficiency to that of a Sharp tip. For a sharp Si tip the SPP excitation is maximum at *d* = 0, i.e. the tip is touching the gold. The efficiency gradually decays as *d* increases. For a 500 nm flat tip the maximum SPP coupling efficiency is achieved at *d* = 100 nm. In an addition, the peak excitation efficiency for a flattened tip (500 nm wide) is lower than that for the case of a sharp tip due to reduced field intensity at the tip region and increased reflection at the flat surface.

From the finding of the above investigations we choose the parameters for the optimum geometry of the presented SPP excitation scheme to be a $54.7^0$ slope angle (*θ*) and a 100 nm distance (*d*) from the flat tip to the Au-SiO$_2$ interface. These values will be used throughout the rest of the paper (unless specified). With the optimum geometry of a sharp tip as discussed above, 29% of the power at 1550 nm wavelength of the incident broad beam at the base of the Si tip (10.4 μm beam diameter) is converted to the characteristic SPP mode of the gold-glass interface on one side of the Si tip at the 1550 nm wavelength. Hence the total power coupled to SPP mode is 58% of the incident power considering SPP excited on both sides of the Si tip. For a 500 nm flat tip setup the single sided coupling efficiency drops down to ~23.56% at 1550 nm as mentioned earlier (~ 47% on both sides). In other words, this design has relatively high tolerance to the sharpness of the Si tip which makes the fabrication of the device less complicated.

Until this point we have discussed about an SPP excitation scheme with a bi-directional excitation. But it is important to be able to excite a directional SPP mode, or a controllability over the excitation direction. Using the same setup as shown in figure 1(a), it is possible to excite a unidirectional SPP by displacing the position of the incoming beam. A displacement towards the right excites an SPP on the left hand side while a displacement of the source to the left excites the SPP on the right side of the tip. The directional excitation of SPP by source displacement is depicted in figure 2(a) and (b). The calculated directional excitation efficiency is ~49% at 1550 nm wavelength for a flat inverted Si tip. Figure 2 (c) shows the extinction ratio of the power of the SPP excited to the left of the tip to that of the right side which is larger than 14 dB for the broad wavelength window of 1500 to 1600 nm.

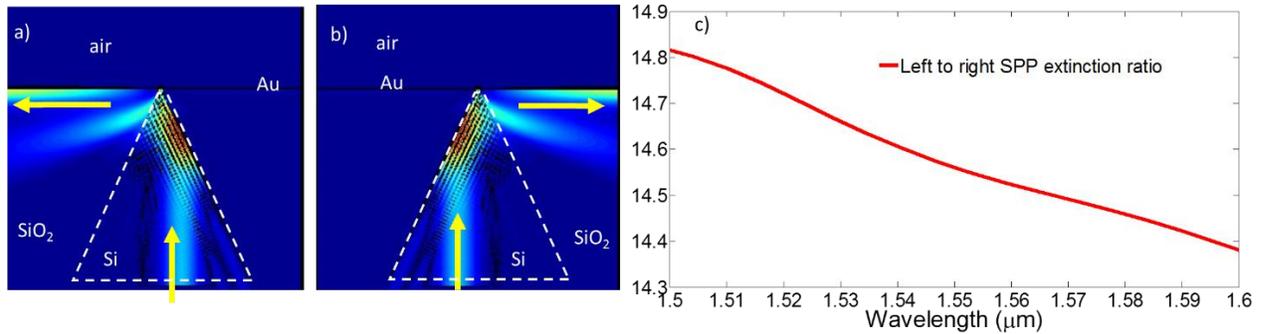

Figure: 2. Power profile for unidirectional SPP excitation on a) the left side by displacing the source to the right, b) the right side by displacing the source to the left. The source was displaced by 1.89 μm from the center of the base of the Si gabled tip. c) Extinction ratio of the power of the SPP excited on the left to that on the right for a source displacement to the right.

To summarize, we have proposed and demonstrated a new very compact technique to efficiently excite an SPP mode at a metal dielectric interface using a high index (Si) gabled tip structure at the 1550 nm wavelength. In our analysis, a perfectly fabricated sharp prism shaped tip with the optimized parameters would excite SPP with 29% efficiency (on each side of the tip) with respect to the power carried by a broad beam of 10.4 μm. For a flattened tip the excitation efficiency can reach a 23.56% (on each side) which is orders of magnitude higher compared to the other compact SPP excitations schemes in literature. A slight displacement of the source from the center of the base excites the SPP in a unidirectional fashion with an efficiency ~ 49% which is higher than any theoretically achievable SPP excitation efficiency reported in literature to our knowledge.